\documentclass[superscriptaddress,twocolumn,showpacs,prb,floatfix]{revtex4}
\usepackage{epsfig}

\begin{document}

\title{Energy size effects of two-dimensional Ising spin glasses}

\author{I.~A.~Campbell}
\affiliation{Laboratoire des Verres, Universit\'e Montpellier II,
34095 Montpellier, France}
\author{Alexander K. Hartmann}
\affiliation{Institut f\"ur Theoretische Physik,
Universit\"at G\"ottingen, Tammannstra\ss{}e 1, 37077 G\"{o}ttingen, Germany}
\author{Helmut G.~Katzgraber}
\affiliation{Theoretische Physik, ETH H\"onggerberg,
CH-8093 Z\"urich, Switzerland}

\date{\today}

\begin{abstract}
We analyze exact ground-state energies of two-dimensional Ising spin
glasses with either Gaussian or bimodal nearest-neighbor interactions 
for large system sizes and for three types of boundary conditions: 
free on both axes, periodic on both axes, and free on one axis and periodic 
on the other. We find accurate values for bulk-, edge-, and corner-site 
energies. Fits for the system with Gaussian bonds are excellent for all 
types of boundary conditions over the whole range of system sizes. In 
particular, the leading behavior for nonfree boundary conditions is 
governed by the stiffness exponent $\theta \approx -0.282$ describing the 
scaling of domain-wall and droplet excitations. For the system with a 
bimodal distribution of bonds the fit is good for free boundary conditions 
but worse for other geometries, particularly for periodic-free boundary 
conditions where there appear to be unorthodox corrections to scaling up 
to large sizes. Finally, by introducing hard bonds we test explicitly for 
the Gaussian case the relationship between domain walls and the standard 
scaling behavior.
\end{abstract}

\maketitle

\section{Introduction}
\label{sec:introduction}

The properties of the two-dimensional Edwards-Anderson Ising spin glass
(ISG)\cite{binder:86,mezard:87,fisher:91c,young:98,kawashima:03}
with either Gaussian or bimodal ($\pm J$) nearest-neighbor interactions have
been intensively studied. Efficient 
algorithms\cite{bieche:80,barahona:82b,derigs:91} have been developed
which provide exact ground-state energies and configurations for these
systems up to large system sizes with $N = L \times L$ spins up 
to\cite{palmer:99} $L = 1800$, allowing for accurate measurements of 
fundamental parameters and
stringent tests of theoretical predictions including precise evaluations
of size-effect corrections to scaling.

In both Gaussian and bimodal disorder distributions, 
freezing occurs only at zero temperature,\cite{hartmann:01a}
i.e.~$T_{\rm c} = 0$.  
For the Gaussian ISG as for any ISG system with a
continuous distribution of interactions there is a unique ground state
(together with its reversed image). At $T = 0$ 
the thermodynamic limit correlation function $G(r)$ is constant, $G(r) = 1$
for all distances $r$. As
the critical exponent $\eta$ is defined through $G(r) \sim
r^{-(d-2+\eta)}$ at the ordering temperature, $\eta=0$. For the $\pm J$
ISG, the ground state is highly degenerate, with a residual entropy at
zero temperature\cite{zhan:00} $S(T = 0)=0.078(5){\it k_{\rm B}}$. In
consequence the time averaged correlation function $G(r)$ drops with 
increasing $r$, meaning
$\eta$ must be positive.\cite{houdayer:01}
The domain-wall stiffness has been measured accurately
through comparisons of the energies of samples with periodic and
antiperiodic boundary conditions in one direction. For the Gaussian ISG,
$\theta = -0.282(2)$.\cite{hartmann:01a,rieger:96} From the scaling rules
applied at a zero temperature transition,\cite{katzgraber:02a,katzgraber:04}
we obtain for the
critical exponent $\nu = -1/\theta$, hence 
$\nu=3.55(3)$.\cite{katzgraber:04,houdayer:04} For the $\pm J$ 2d
ISG, large-$L$ simulations\cite{hartmann:01a} 
on samples with periodic/antiperiodic boundary
conditions along one axis and free boundary conditions on the other axis
showed $\theta=0$, with significant corrections to scaling up to $L \sim
100$. Finite-size scaling and Migdal-Kadanoff
simulations on a variety of 2d ISG samples with different distributions of
interactions are consistent with a unique lower critical dimension $d_l
\sim 2.5$; for all the samples with discrete distributions of interactions
$\theta=0$ at all dimensions below the lower critical 
dimension.\cite{amoruso:03} For these systems the effective exponent $\nu$ 
is infinite. 
The correlation length diverges exponentially \cite{houdayer:01} and the
specific heat drops to zero as $\exp(-n\beta)$; data were analyzed using $n$
equal to\cite{saul:93,saul:94} $4$ but recent work\cite{lukic:03}
is consistent with
$n=2$. Although we will be concerned here only with 
zero-temperature behavior, we can note that there have been conflicting
estimates of the ``droplet'' excitation exponent for the Gaussian ISG, which
standard arguments indicate should be identical to $\theta$. It has been
suggested that apparent disagreements are artifacts due to corrections to
scaling\cite{moore:02} but Berthier and Young 
found\cite{berthier:03} that whether such corrections are visible 
seems to depend on the details of the measurements.

We have carried out measurements of the ground-state energy per spin on
square samples as a function of system size $L$ for three different types of
boundary condition: free along both $x$ and $y$ pairs of boundaries (to
be referred to as ffbc), free along one pair of boundaries and periodic
along the other pair to give a cylindrical geometry (pfbc), and periodic
along both pairs of boundaries to give a toroidal geometry (ppbc).  These
three geometries are physically distinct. In the ffbc case, there are bulk
sites, edge sites, and corner sites. However the system has
no boundary condition constraints, or in other words it can be taken to
contain no ``domain walls,'' at least for a sample with a unique ground
state. In the pfbc scenario there are edge sites but no 
corner sites; on the other
hand the sample is constrained by the boundary condition along one
direction. Finally for the familiar ppbc case there are no edges or corners but
there are constraints along both directions. It is for this geometry with
no boundaries that the standard scaling approach should apply.

The behavior of the ffbc systems can be explained
in simple geometrical terms. The behavior of the Gaussian systems with
pfbc and ppbc is governed by an algebraic term with the exponent 
$\theta \approx -0.282$, the same value as the 
exponent describing the behavior of domain-wall
and droplet excitations.\cite{hartmann:02c,hartmann:03}
For this reason, we also compare sample by sample ffbc and pfbc 
realizations with the same interaction set and look for the appearance of 
system-spanning domain walls. Finally, we introduce 
{\em hard bonds} to force system-spanning domain walls and study
how the appearance of these depends on the fraction of hard bonds.
  
The paper is organized as follows:
In Sec.~\ref{sec:algorithms} we introduce the model, observables, 
and details of the algorithms used. Scaling arguments are 
summarized in Sec.~\ref{sec:scaling}.
Results for the ffbc, ppbc, and pfbc scenarios are presented in
Secs.~\ref{sec:ffbc}, \ref{sec:ppbc}, and \ref{sec:pfbc}, respectively.
Domain-wall calculations are discussed in Sec.~\ref{sec:dwp} and concluding
remarks are contained in Sec.~\ref{sec:conclusions}.

\section{Model and Algorithms}
\label{sec:algorithms}

The Hamiltonian of the two-dimensional Ising spin glass is given by
\begin{equation}
{\cal H} = -\sum_{\langle i,j \rangle} J_{ij} S_i S_j ,
\label{eq:hamiltonian}
\end{equation}
where the sites $i$ lie on a square lattice in two dimensions and
the $J_{ij}$ are nearest-neighbor interactions. In the Gaussian case the
couplings
are chosen according to a Gaussian distribution with zero mean and
standard deviation unity, whereas in the bimodal case the couplings
$J_{ij}$ can take values of $\{\pm 1\}$ with equal probability.

In greater than two dimensions, or in the presence of a magnetic field,
the exact calculation of spin-glass ground states belongs to the class of
$NP$-hard problems.\cite{barahona:82,hartmann:02d} This means that only
algorithms with exponentially increasing running times are known.  Here,
we have studied mainly square lattices with periodic boundary conditions
in at most one direction without external magnetic fields. For this
special case of a planar system there are efficient polynomial-time
``matching'' algorithms.\cite{bieche:80} The basic idea is to represent
each realization of the disorder by its frustrated 
plaquettes.\cite{toulouse:77}
Pairs of frustrated plaquettes are connected by paths
in the lattice and the weight of a path is defined by the sum of the
absolute values of the coupling constants which are crossed by the path. A
ground state corresponds to the set of paths with minimum total weight,
such that each frustrated plaquette is connected to exactly one other
frustrated plaquette. This is called a minimum-weight perfect matching.
The bonds which are crossed by paths connecting the frustrated plaquettes
are unsatisfied in the ground state, and all other bonds are satisfied.

For the calculation of the minimum-weight perfect matching, efficient
polynomial-time algorithms are available.\cite{barahona:82b,derigs:91}
Recently, an implementation has been presented,\cite{palmer:99} where
ground-state energies of large systems of size $N\le 1800^2$ were
calculated.  Here, an algorithm from the LEDA library\cite{mehlhorn:99}
has been applied, which limits the system sizes due to the restricted size
of the main memory of the computers which we used.

Furthermore, we have studied two-dimensional systems with fully periodic
boundary conditions. Here, we use the spin-glass ground-state server
at the University of Cologne\cite{juenger:sg} where a
``branch-and-cut'' method\cite{desimone:95,desimone:96} (see
Ref.~\onlinecite{juenger:01} for a tutorial on optimization problems and
techniques, including branch-and-bound and branch-and-cut) is used, which is
currently the fastest exact algorithm for computing spin-glass ground
states,\cite{palassini:02} with the exception of the polynomial-time
special cases mentioned above. Nevertheless, the CPU time increases faster
than a power of the system size. Therefore we cannot study
as large systems as are possible with the matching algorithm.
Nonetheless, the implementation of the branch-and-cut algorithm on the
Cologne server is very efficient, hence we can still study quite a large
range of sizes, in practice $L \le 64$.

The numbers of samples over which averages are taken are shown in
Tables \ref{tab:paramsgsbc} and 
\ref{tab:paramsppbc}.

\begin{table}
\caption{
The number of samples for the ground-state calculations for
different system sizes, distributions and free-free (ffbc), 
resp.,~free-periodic (pfbc) boundary conditions.
\label{tab:paramsgsbc}
}
\begin{tabular*}{\columnwidth}{@{\extracolsep{\fill}} c l l l l  }
\hline
\hline
$L$ & Gaussian ffbc    & Gaussian pfbc    & $\pm J$ ffbc   & $\pm J$ pfbc    \\
\hline
2   & $1\times 10^6$   &                  & $4\times 10^6$ &                 \\
3   & $1\times 10^6$   & $1\times 10^6$   & $1\times 10^6$ & $1\times 10^6$  \\
4   & $9.1\times 10^5$ & $1\times 10^6$   & $1\times 10^6$ & $9.4\times 10^5$\\
6   & $5.1\times 10^5$ & $9.9\times 10^4$ & $5\times10^5$  &$5.4 \times 10^5$\\
8   & $2.1\times 10^5$ & $9.9\times 10^4$ & $2\times 10^5$ & $3.4\times 10^5$\\
10  & $1.1\times 10^5$ & $9.9\times 10^4$ & $1\times 10^5$ & $1.4\times 10^5$\\
16  & $4\times 10^4$   & $9.9\times 10^4$ & $1\times 10^4$ & $4\times 10^4$  \\
20  & $1\times 10^4$   & $1\times 10^4$   & $1\times 10^4$ & $4\times 10^4$  \\
30  & $1\times 10^4$   & $1\times 10^4$   & $1\times 10^4$ & $4\times 10^4$  \\
40  & $1\times 10^4$   & $1\times 10^4$   & $1\times 10^4$ & $3.9\times 10^4$\\
60  & $1\times 10^4$   & $1\times 10^4$   & $1\times 10^4$ & $4\times 10^4$  \\
80  & $1\times 10^4$   & $1\times 10^4$   & $1\times 10^4$ & $4\times 10^4$  \\
120 & $1\times 10^4$   & $1\times 10^4$   & $1\times 10^4$ & $4\times 10^4$  \\
160 & $1\times 10^4$   & $1.1\times 10^4$ & $1\times 10^4$ & $4\times 10^4$  \\
240 & $724$            & $2800$           & $5\times 10^5$ & $4.6\times 10^4$\\
320 &                  & $3710$           & $5\times 10^5$ & $2.4\times 10^4$\\
480 &                  &                  &                & $2\times 10^4$  \\
\hline
\hline
\end{tabular*}
\end{table}

\begin{table}
\caption{
The number of samples for the ground-state calculations for full periodic 
(ppbc) boundary conditions, for different system sizes and disorder 
distributions.
\label{tab:paramsppbc}
}
\begin{tabular*}{\columnwidth}{@{\extracolsep{\fill}} c c l  }
\hline
\hline
$L$  &  Gaussian & $\pm J$ \\
\hline
3  & $5\times 10^4$ & $1\times 10^4$ \\
4  & $2\times 10^4$ & $1\times 10^4$ \\
6  & $1\times 10^4$ & $5\times 10^3$ \\
8  & $1\times 10^4$ & $5\times 10^3$ \\
12 & $5\times 10^3$ & $2\times 10^3$ \\
16 & $1\times 10^3$ & $2\times 10^3$ \\
24 &                & $2\times 10^3$ \\
32 & $2\times 10^3$ & $2\times 10^3$ \\
64 & $1\times 10^3$ &                \\
\hline
\hline
\end{tabular*}
\end{table}

\section{Scaling}
\label{sec:scaling}

We want to find the size-dependence of the energy from simple scaling
arguments for the generic ppbc case and follow closely along the lines 
of the scaling discussion presented in Ref.~\onlinecite{katzgraber:02a}. 
The internal energy per spin $e$ of 
a thermodynamic system can be expressed as a derivative of the free 
energy per spin $f$ with respect to the temperature:
\begin{equation}
e = -\frac{1}{\beta^2}\frac{d(\beta f)}{dT} \; ,
\end{equation}
where $\beta=1/T$, $T$ the temperature.

We assume that
the scaling of the temperature dependence of the correlation length is
$\xi \sim (T-T_{\rm c})^{-\nu}$ or $(T-T_{\rm c}) \sim \xi^{-1/\nu}$.
Therefore
\begin{equation}
dT \sim \xi^{-(1+1/\nu)} d\xi \; .
\end{equation}
Hence, we obtain
\begin{equation}
e(T) = - \frac{1}{\beta^2} \frac{d(\beta f)}{d\xi}\frac{d\xi}{dT} + B(T) \; .
\end{equation}
$B(T)$ is the smooth noncritical background.

Using basic finite-size scaling arguments,\cite{yeomans:92} we know that for the singular part of the free energy $\beta f  
\sim \xi^{-d}$.\cite{katzgraber:02}

First we discuss the case $T_{\rm c} > 0$. In this case the factor $1/\beta^2$
is not critical. Therefore we obtain
\begin{equation}
e(T) \sim \frac{1}{\beta_{\rm c}^2}\xi^{-(d-1/\nu)} + B(T) \; ,
\end{equation}
where $\beta_{\rm c} = 1/T_{\rm c}$.
At $T_{\rm c}$,  $ \xi_L = {\mathcal C}_1 L$, ${\mathcal C}_1$
a constant, so the critical behavior  of $e_L$ is
\begin{equation}
e_L -e_{\infty} \sim L^{-(d-1/\nu)} \;\;\;\;\;\;\;\;\;\;(T_{\rm c} > 0) \; .
\label{eq:e}
\end{equation}

Next, we discuss the case $T_{\rm c} = 0$.
Now $\beta$ is critical, $\beta \sim \xi^{1/\nu}$, hence
\begin{equation}
e  \sim \frac{1}{\xi^{2/\nu}} \xi^{-(d-1/\nu)} + B =
\xi^{-(d+1/\nu)}  + B \; .
\end{equation}
For a finite system of size $L$,
\begin{equation}
e_L - e_{\infty} \sim  L^{-(d+1/\nu)} \;\;\;\;\;\;\;\;\;\;(T_{\rm c} = 0) \; .
\label{eq:scale_T0}
\end{equation}

For $T_{\rm c} = 0$ one also has the scaling relation\cite{bray:84}
between the domain-wall
stiffness exponent $\theta$ and the exponent $\nu$,  $\theta = -1/\nu$,
therefore
\begin{equation}
e_L -e_\infty \sim  L^{-(d-\theta)} \;\;\;\;\;\;\;\;\;\;(T_{\rm c} = 0) \; .
\label{eq:scale_T0b}
\end{equation}

For the $\pm J$ case, $T_c=0$ and the correlation length diverges
exponentially. Following through the algebra with\cite{houdayer:01} 
$\xi = \exp(2/\beta)$
leads to $e_L -e_\infty \sim  L^{-d}$, which is just the same as
Eq.~(\ref{eq:scale_T0b}) with $\theta = 0$, corresponding to $\nu=\infty$.

Equation (\ref{eq:scale_T0}) is a scaling identity for any system with 
ppbc and $T_{\rm c}=0$, and there is no need to appeal to specific physical 
arguments to justify it.  Any deviations from this identity must be due to
corrections to scaling. (We can note that Bouchaud {\em et
  al}.\cite{bouchaud:02} refer to this  scaling term as the
``correction to scaling.'') 

The leading correction to scaling should be either a renormalization group
theory (RGT) irrelevant operator correction giving the form
\begin{equation}
  e_L-e_{\infty} \sim L^{-(d-\theta)}(1+{\mathcal K}_{1}L^{-\omega} + \cdots)
\end{equation}
where $\omega$ is the RGT leading correction to scaling exponent and
${\mathcal K}_{1}$ is a constant, or an analytic correction which 
plausibly introduces a term  ${\mathcal K}_{2}L^{-2}$.\cite{salas:99,salas:00}
This ``analytic'' term is unrelated to the RGT irrelevant operators. 
The arguments leading to a prediction for the energy correction at this 
level can be very involved, even for the canonical $2d$ Ising 
ferromagnet where there is no irrelevant operator term but there are analytic terms in odd powers of $1/L$.\cite{salas:99,salas:00,salas:01}
For ISGs the leading term in the RGT $\epsilon$-expansion for $\omega(d)$
as a function of dimension $d$ is\cite{dedominicis:03}: $\omega(d)=
(6-d) + \cdots$; as we can safely assume that by $d = 2$, 
$\theta + \omega \gg 2$ hence the dominating correction term is an 
analytic term.

Thus finally, for ppbc from standard scaling arguments we expect
\begin{equation}
e_L = e_{\infty} + {\mathcal L}_{1}L^{-(d-\theta)}  
+ {\mathcal L}_{2}L^{-2} + \cdots \; ,
\end{equation}
where the ${\mathcal L}_{1}$ term is the scaling term and the 
${\mathcal L}_{2}$
term is the analytic correction term. In practice, if $\theta$ is small
compared to $d$, it will be hard to distinguish between this sum of two
terms and an ``effective'' scaling,
\begin{equation}
e_L = e_{\infty} + {\mathcal L}_{\rm eff}L^{-(d-\theta_{\rm eff})} \; ,
\end{equation}
with a single effective exponent $\theta_{\rm eff}$ whose value is a
function of ${\mathcal L}_{2}/{\mathcal L}_{1}$ and which can be larger or 
smaller than $\theta$ depending on the sign of this ratio.

Although the remark is irrelevant for the two-dimensional case, we can note
that if $T_{\rm c} > 0$ this discussion is valid {\it mutatis mutandi} at 
$T_{\rm c}$; however if $T_{\rm c} > 0$ scaling rules do not apply 
below $T_{\rm c}$ and in particular at $T=0$. Alternative physical arguments
must be used for discussing the ground-state behavior.

\section{Free boundary conditions (ffbc)}
\label{sec:ffbc}

If boundary conditions are free along all four edges we can
consider that there are no external constraints on the system. There are
however edge and corner sites, edge bonds and corner bonds. The total
number of bonds is equal to $2(L^{2}-L)$ while the number of spins is
$L^2$, hence we present a discussion in terms of the energy per spin 
$e_L$ or alternatively twice the energy per bond $2e_L^{\rm bond}$ (the values will become identical in the infinite $L$ limit). 

\begin{figure}
\centerline{\epsfxsize=\columnwidth \epsfbox{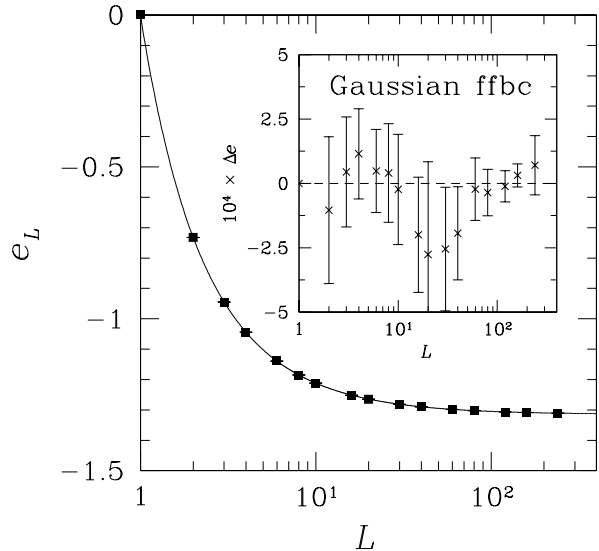}}
\vspace{-1.0cm}
\caption{
Energy per spin for the ffbc Gaussian case. The solid line represents a 
three-parameter fit to Eq.~(\ref{eq:energyffbc}) with $\chi^2 = 0.67$.
The inset shows the difference between
the fit and the actual data, $\Delta e$, represented as an energy per spin.
Fit parameters are summarized in Table \ref{tab:fitparams_ffbc}.
}
\label{fig:gauss_ffbc_fig}
\end{figure}

A heuristic geometrical
scaling consists in writing twice the energy per bond as
\begin{equation}
2e_L^{\rm bond}= {\mathcal A}^{*} + {\mathcal B}^{*}/L + 
{\mathcal C}^{*}/L^2 \; ,
\label{eq:gscale}
\end{equation}
where ${\mathcal A}^{*}$, ${\mathcal B}^{*}$, and ${\mathcal C}^{*}$
are constants.
Physically the first term in Eq.~(\ref{eq:gscale}) represents the 
energy per bulk bond, the
second term is a correction for the energy difference between an edge
and a bulk bond, and the third term a further correction for a corner
bond. Even if the localization of the energy differences onto the edge
or corner bonds is only approximate so the identification between terms and
bonds is not rigorous, on geometrical grounds the ffbc energy size effects
can be expected to have strictly this form with no further terms, at least down to
small $L$ values where neighboring corner effects begin to interfere.
Translating Eq.~(\ref{eq:gscale}) into terms of energy per spin,
i.e. using $E_L=(L^2-L)e_L^{\rm bond}=L^2e_L$, gives an energy per spin,
\begin{equation}
e_L = {\mathcal A}^{*} -({\mathcal A}^{*} - {\mathcal B}^{*})/L 
- ({\mathcal B}^{*} - {\mathcal C}^{*})/L^2 - {\mathcal C}^{*}/L^3 \; .
\label{eq:energyffbc}
\end{equation}
For the isolated spin ($L=1$) the energy per bond is not defined, but the
energy per spin is identically zero. By inspection Eq.~(\ref{eq:energyffbc}) 
happens to have a
form such that $e_L$ must be exactly equal to zero for $L=1$. We can
also write
\begin{equation}
e_L= {\mathcal A} + {\mathcal B}/L + {\mathcal C}/L^2 + {\mathcal D}/L^3 \; ,
\label{eq:energy2}
\end{equation}
with ${\mathcal A} = {\mathcal A}^{*}$, 
${\mathcal B} = - ({\mathcal A}^{*} - {\mathcal B}^{*})$, 
${\mathcal C} = - ({\mathcal B}^{*} - {\mathcal C}^{*})$, 
and ${\mathcal D} = - {\mathcal C}^{*}$.
For the Gaussian ffbc data the fit to Eq.~(\ref{eq:energy2})
is shown in Fig.~\ref{fig:gauss_ffbc_fig}. It can be seen that for
the entire range of sizes from $L=1$ to $L=240$ the three-parameter fit is
excellent with $\chi^2 = 0.67$. 
The same data with the same fit parameters ${\mathcal A}$, ${\mathcal B}$, 
${\mathcal C}$, and ${\mathcal D}$ are presented as a difference plot,
i.e. as the difference between the fit and the actual data, in 
the inset of Fig.~\ref{fig:gauss_ffbc_fig}.
Remarkably, this naive equation with three free parameters is in agreement 
with the measurements to within one or two parts in $10^4$ for the whole 
range of $L$ from $L=240$ down to and including $L=1$. The fit parameters are
given in Table \ref{tab:fitparams_ffbc}.

\begin{table}
\caption{
Fit parameters according to Eq.~(\ref{eq:energyffbc}) 
for ff boundary conditions, as well as for Gaussian and bimodal distributed
disorder.
\label{tab:fitparams_ffbc}
}
\begin{tabular*}{\columnwidth}{@{\extracolsep{\fill}} c l l  }
\hline
\hline
parameter  &  Gaussian & $\pm J$ \\
\hline
${\mathcal A}^{*}$   & $-1.31479(2)$ & $-1.40197(2)$ \\
${\mathcal B}^{*}$   & $-0.3205(9)$  & $-0.5492(20)$ \\
${\mathcal C}^{*}$   & $0.042(3)$    & $0.506(18)$   \\
\hline
\hline
\end{tabular*}
\end{table}

\begin{figure}
\centerline{\epsfxsize=\columnwidth \epsfbox{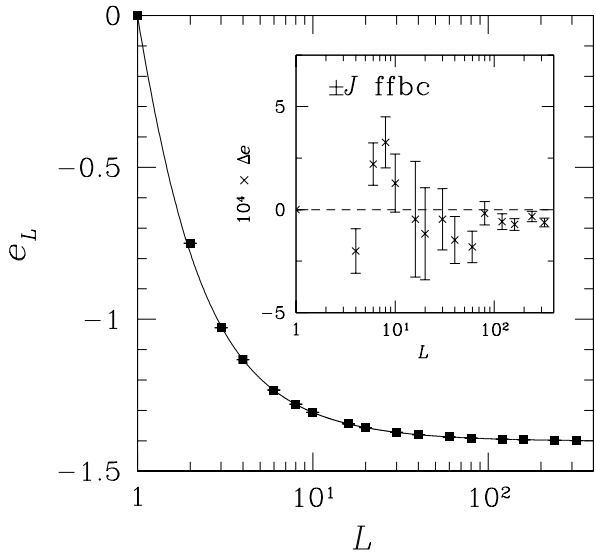}}
\vspace{-1.0cm}
\caption{
Energy per bond for the ffbc $\pm J$ case. The solid line corresponds to a
three-parameter fit to Eq.~(\ref{eq:energyffbc}) with $\chi^2 = 1.29$.
The inset shows the difference between
the fit and the actual data.
Fit parameters are summarized in Table \ref{tab:fitparams_ffbc}.
}
\label{fig:pm_ffbc_fig}
\end{figure}

If we repeat the same analysis for $\pm J$ bonds, using data from $L = 4$ to
$L = 320$, the agreement is almost equally good, cf.~Fig.~\ref{fig:pm_ffbc_fig}.
The fit parameters ($\chi^2=1.29$) 
are given in the Table \ref{tab:fitparams_ffbc}. 
(We have not included in the fit the point for
$L = 2$ which lies almost $3\%$ above the trend. By this size each system
has only four bonds and there are only five possible energies for any $\pm J$
system.  As one might foresee, quantization effects seem to break down
the precise scaling by this size.)

The fit extrapolates to a bulk (or infinite lattice size) energy per spin of
\begin{equation}
e_{\rm bulk}^{\rm Gauss} = -1.31479(2) \; ,
\end{equation}
for the Gaussian case. The most recent published value is 
$e_{\rm bulk}^{\rm Gauss} = -1.317(1)$\cite{desimone:95}.
For the $\pm J$ case the present data give a bulk energy per spin of
\begin{equation}
e_{\rm bulk}^{\pm J} = -1.40197(2) \; .
\end{equation}
This estimate is compatible with the very precise published value 
$e_{\rm bulk}^{\pm J} = -1.401938(2)$.\cite{palmer:99}
Notice that the bulk energy of the bimodal system is significantly more
negative that the bulk energy of the Gaussian system. 

In the ffbc lattice with $L^2$ spins there are $(L-2)^2$ bulk sites, $4L-8$
edge sites, and $4$ corner sites. We associate energies $e_{\rm bulk}$,
$e_{\rm edge}$, and $e_{\rm corner}$ with sites at each of these positions, 
respectively (equal to half the sum of the adjacent bond energies).
Using $E_L=L^2e_L=(L-2)^2e_{\rm bulk}+(4L-8)e_{\rm edge}+4e_{\rm corner}$ 
and comparing with Eq.~(\ref{eq:energy2}), we can  make the identifications 
for the energies per site,
\begin{eqnarray}
\label{eq:bond-energies}
e_{\rm bulk} & = & {\mathcal A}\; ,\nonumber\\
e_{\rm edge} & = & {\mathcal A} + {\mathcal B}/4\; ,\\\nonumber 
e_{\rm corner} & = & {\mathcal A} + {\mathcal B}/2 + {\mathcal C}/4 \; .
\end{eqnarray} 
Because each bulk site has four bonds, each edge site three bonds, and each
corner site two bonds, and because each bond is 
shared with a neighbor site, to obtain
the average bond energies for the bonds linking to these three 
types of sites we divide $e_{\rm bulk}$, $e_{\rm edge}$, $e_{\rm corner}$ 
by $2$, $1.5$, and $1$, respectively.

The bulk, edge, and corner energies per bond evaluated according to 
Eq.~(\ref{eq:bond-energies}) when inserting the measured scaling parameters 
${\mathcal A}$, ${\mathcal B}$, and ${\mathcal C}$ are shown in Table
\ref{tab:eperbond_ffbc}.
As is to be expected in both cases the edge 
and corner energies are more negative than the bulk bond energies because 
the bonds at the edges and corners are less frustrated. The de-frustration 
effects can be seen to be $\sim 50\%$ stronger in the $\pm J$ case as 
compared to the system with Gaussian interactions.

\begin{table}
\caption{
Energy per bond for ff boundary conditions. The edge and corner energies are
larger in magnitude than the bulk values because the bonds at the edges and
corners are less frustrated.
\label{tab:eperbond_ffbc}
}
\begin{tabular*}{\columnwidth}{@{\extracolsep{\fill}} c l l  }
\hline
\hline
bond  &  Gaussian & $\pm J$ \\
\hline
Bulk  & $-0.65740(1)$ & $-0.70099(1)$ \\
Edge  & $-0.7108(2)$ & $-0.7925(4)$ \\
Corner  & $-0.727(1)$ & $-0.821(4)$ \\
\hline
\hline
\end{tabular*}
\end{table}

The Gaussian values can be compared with explicit bond-by-bond 
sample-by-sample measurements of bulk, edge, and corner bond energies. 
Averaging over 100 $L=100$ ground states with Gaussian bonds we obtain 
$-0.6576(4)$, $-0.698(2)$, and $-0.66(2)$ for the mean local bulk, edge 
and corner bond energies, respectively. These values are compatible with 
the more precise values estimated by scaling, allowing for the fact that 
the definitions are identical for the bulk bonds but not for the edge or 
corner bonds. In these cases the scaling values implicitly include energy 
changes at slightly perturbed further bonds.

\section{Periodic-Periodic boundary conditions (ppbc)}
\label{sec:ppbc}

Periodic-periodic boundary conditions are the standard geometry in which 
simulations are conventionally carried out, and to which the critical 
scaling rules discussed in Sec.~\ref{sec:scaling} [Eq.~(\ref{eq:scale_T0})]
should apply.  
There are no edges or corners and all sites
are on average in equivalent environments. The infinite-size limit energy
per spin should be identical to the infinite size bulk energy per spin in
the ffbc geometry.

\begin{figure}
\centerline{\epsfxsize=\columnwidth \epsfbox{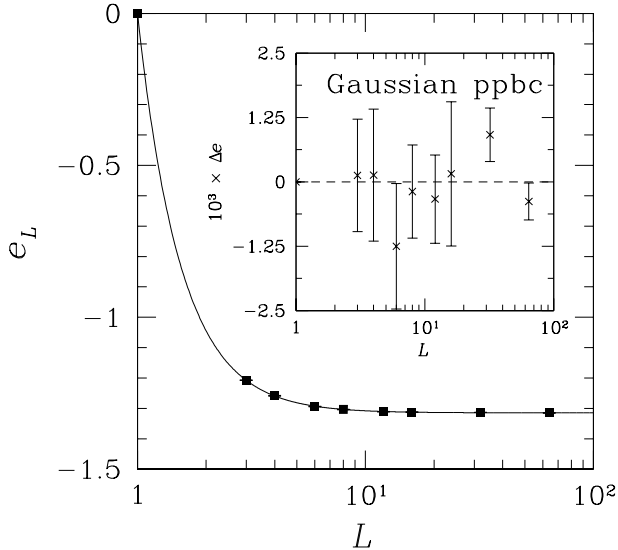}}
\vspace{-1.0cm}
\caption{
Energy for the ppbc Gaussian case. The solid line corresponds to a 
``fit'' with {\em no free parameters} to Eq.~(\ref{eq:scale_ex})  
which yields a quality of fit
$\chi^2 = 0.92$. The inset shows the difference between
the fit and the actual data.
}
\label{fig:gauss_ppbc_fig}
\end{figure}

For the Gaussian case, high precision domain-wall measurements give
excellent scaling with an exponent 
$\theta=-0.282(2)$,\cite{hartmann:01a,rieger:96} and we should expect
\begin{equation}
e_L - e_{\infty} \sim L^{-(2-\theta)} \; ,
\end{equation}
up to correction terms. In fact we go further and test much more
stringent assumptions: first, that corrections are negligible, and
second that the energy per spin at large and moderate $L$ extrapolates
exactly to $e_L = 0$ at $L=1$. Thus we test the relation
\begin{equation}
e_L = e_{\infty}\left[1-L^{-(2-\theta)}\right] \; .
\label{eq:scale_ex}
\end{equation}
Here we have no free parameters as we adopt the value
$e_{\infty} = -1.31479$ obtained above in the ffbc measurement, and the
value $\theta = -0.282$ obtained from domain-wall measurements. We exclude
$L=2$ as this size is pathological in ppbc because of wrap-around effects
for the interactions. For the ``fit'' $\chi^2 = 0.92$ for sizes from $L=3$ to
$64$. The data are represented in Fig.~\ref{fig:gauss_ppbc_fig}
with a difference
plot in the inset. It can be seen that this parameter-free expression
represents the high precision data to within the statistical accuracy.
Alternatively, assuming that the $e_{\infty}$ value is correct and that
there are no corrections to scaling, we can fit leaving $\theta$ as a free
parameter. We obtain a best fit for the stiffness exponent,
\begin{equation}
\theta = -0.281(7) \; ,
\end{equation}
i.e.,~$\nu \approx 3.55$, with $\chi^2 = 0.78$. This is an accurate
independent value in excellent agreement with the domain-wall estimates,
as well as finite-temperature estimates.\cite{katzgraber:04,houdayer:04}
The assumptions that we have made are either exact or are very good
approximations.

\begin{figure}
\centerline{\epsfxsize=\columnwidth \epsfbox{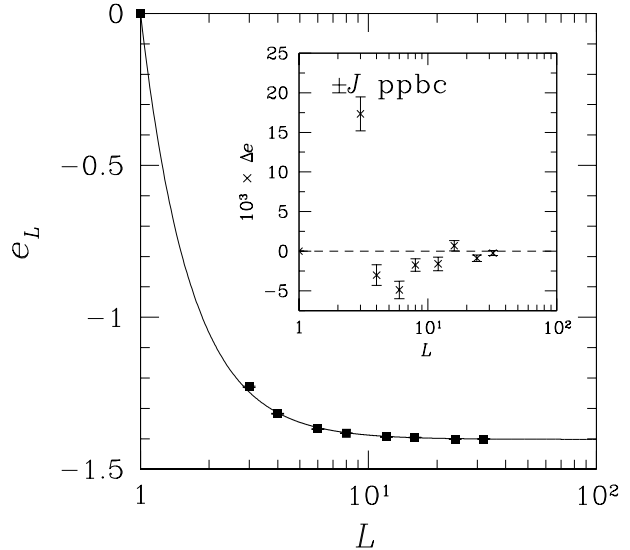}}
\vspace{-1.0cm}
\caption{
Energy for the ppbc $\pm J$ case. The solid line corresponds to a
parameter-free ``fit'' according to Eq.~(\ref{eq:scale_ex}) with 
$e_{\infty}=-1.401938$ and $\theta=0$.
The inset shows the differences between the fit and the actual data. 
}
\label{fig:pm_ppbc_fig}
\end{figure}

For the $\pm J$ case (Fig.~\ref{fig:pm_ppbc_fig}) the situation is less 
clear cut. If we make the same
stringent parameter-free assumptions, taking $e_{\infty}=-1.401938$ and
$\theta=0$ we obtain a rather mediocre fit, $\chi^2 \approx 15$ for the data
from $L=3$ to $L=32$ or $\chi^2 \approx 6.5$ for the data from $L=4$ to
$L=32$. Relaxing the condition $e_{L=1}=0$ gives a better fit, $e_L =
e_{\infty}+1.319/L^2$ 
with $\chi^2 \approx 2.6$ for $L=4$ to $L=32$. This
could be understood as an analytic correction term appearing. Such a term
should be expected to also have the form $L^{-2}$ and so in the $\pm J$
case where $\theta=0$ it would simply modify the prefactor of the scaling
term which is also proportional to $L^{-2}$. Alternatively, if we assume
that for the range of $L$ over which the energy differences can be
measured the {\em effective} value of $\theta$ is different from zero,
i.e., when incorporating the corrections to scaling into it, we can
fit to Eq.~(\ref{eq:scale_ex})
with $\theta$ as a free parameter. The best fit in this case from $L=4$ to
$L=32$ corresponds to $\theta = -0.04(1)$, with $\chi^2 \approx 2.0$.
This discussion indicates that the strict zero free parameter scaling that
gives virtually perfect agreement in the Gaussian case explains the form
of $e_L$ to much lower precision in $\pm J$ case. This could mean that
unknown higher-order corrections to scaling must be taken into 
account to represent the behavior also at the relatively small sizes 
accessible for the ppbc case. We believe that this behavior is due to
the discreteness of the bond distribution, leading to a high ground-state
degeneracy of the $\pm J$ model; see the discussion at the end of the
next section.

\section{Periodic-Free boundary conditions (pfbc)}
\label{sec:pfbc}

We study a cylindrical geometry where the boundary 
conditions are free along one pair of edges, and periodic
for the perpendicular pair. In the large-$L$
limit the bulk site or bond energy term per spin or per bond should again
be exactly the same as in the ffbc case. We assume that the free-edge size
effect per edge site also has the same value as in the ffbc case. There are no
corners. Because the periodic boundary condition imposes a constraint,
{\it a priori} one can also expect terms of the same type as in ppbc.
The edge effects being allowed for, we assume that the scaling term
is again proportional to $L^{-(2-\theta)}$ and we allow for an
analytical correction term proportional to $L^{-2}$. The total number of
bonds is $2L^2-L$. Following the same arguments as in the ffbc 
case presented in Sec.~\ref{sec:ffbc} we test the expression
\begin{equation}
e_L = {\mathcal A}^{*} + \frac{1}{2L}({\mathcal B}^{*} - {\mathcal A}^{*}) 
- \frac{1}{4L^2}{\mathcal B}^{*} + {\mathcal L}_{1}L^{-(2-\theta)}
+ \frac{{\mathcal L}_{2}}{L^2} \; ,
\label{eq:pfbcfit}
\end{equation}
where ${\mathcal A}^{*}$ and ${\mathcal B}^{*}$ are taken directly from the 
ffbc analysis, so we have only two free parameters for the fit.

\begin{figure}
\centerline{\epsfxsize=\columnwidth \epsfbox{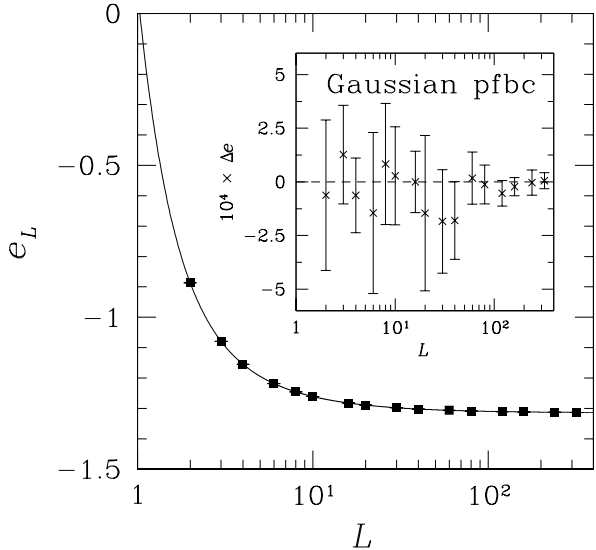}}
\vspace{-1.0cm}
\caption{
Energy for the pfbc Gaussian case and fit to Eq.~(\ref{eq:pfbcfit}) (see 
the text). The inset shows the differences between
the fit and the actual data.
} 
\label{fig:gauss_pfbc_fig}
\end{figure}

For the Gaussian pfbc data, the fit to points from $L=3$ to $L=320$ is
excellent, $\chi^2 = 0.26$ for ${\mathcal L}_{1}= 1.019(9)$ and 
${\mathcal L}_{2}= -0.203(7)$. 
The data are shown in Fig.~\ref{fig:gauss_pfbc_fig}. If there
is no $L^{-2}$ correction term, the fit is significantly worse. We
note that for the pfbc geometry an analytic correction term is
necessary, and that the fit does not extrapolate exactly to $e_L=0$ at
$L=1$.  The fit is thus less aesthetically pleasing than was the case for
either ffbc or the ppbc cases; we can conclude that nevertheless a high quality
consistent analysis can be made of the pfbc Gaussian data over the entire
range of $L$ using physically reasonable assumptions for the fits. However,
it is not clear to us why an analytic correction term is needed for the
pfbc geometry while the ppbc fit is excellent without any such correction.

\begin{figure}
\centerline{\epsfxsize=\columnwidth \epsfbox{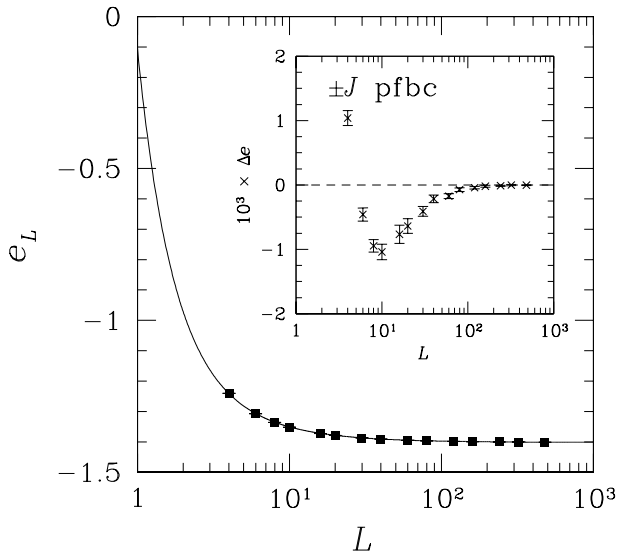}}
\vspace{-1.0cm}
\caption{
Energy for the pfbc $\pm J$ case and fit to Eq.~(\ref{eq:pfbcfit}) (see the 
text). The inset shows the differences between
the fit and the actual data.
}
\label{fig:pm_pfbc_fig}
\end{figure}

For the $\pm J$ data, see Fig.~\ref{fig:pm_pfbc_fig},
we proceed in exactly the same way. In this case the
large-$L$ stiffness exponent\cite{hartmann:01a} $\theta = 0$, so one
should expect both the scaling and the analytical correction to contribute
to a single composite term in $L^{-2}$.  With this assumption the fit to
the pfbc data from $L=3$ to $L=480$ is very poor ($\chi^{2} \approx 30$).

\begin{figure}
\centerline{\epsfxsize=\columnwidth \epsfbox{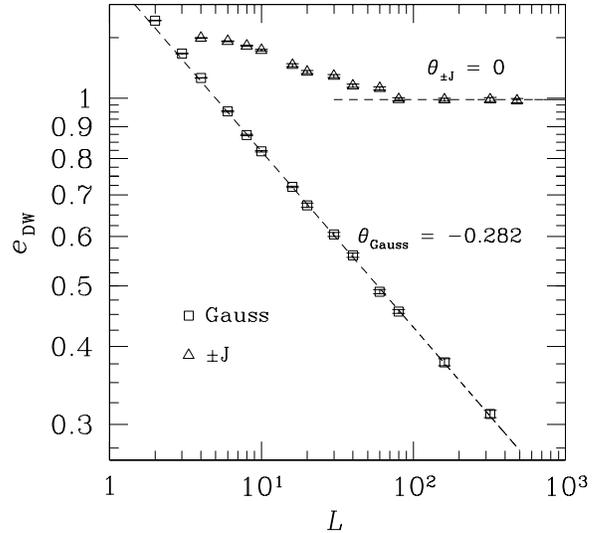}}
\vspace{-1.0cm}
\caption{
Domain-wall energies for Gaussian as well as $\pm J$ bonds. 
While in the Gaussian case the asymptotic power-law behavior appears for
small system sizes, in the $\pm J$ case the asymptotic behavior where $\theta
= 0$ does not appear until lattice sizes $L \approx 100$.
}
\label{fig:dw_energy}
\end{figure}

To understand this behavior, we can note that the domain-wall stiffness
measurements\cite{hartmann:01a} on the $\pm J$ ISG
taken in the same pfbc conditions show strong and nonorthodox corrections
to scaling which extend to $L \sim 100$; the data only
attain the limiting behavior $\theta = 0$ for very large $L$,
as can be seen in Fig.~\ref{fig:dw_energy}. In the
Gaussian case both the domain-wall stiffness and the size dependence of
the energy in the pfbc geometry follow consistently the expected scaling
rules. For the $\pm J$ case in pfbc geometry on the other hand corrections
to scaling of the standard form are not adequate to explain the deviations
from scaling which exist up to large values of $L$. This is probably
due to the discreteness of the interactions for the $\pm J$ case. This
discreteness leads to a high
degeneracy, which in turn allows in two dimensions to 
form domain-walls with almost
no energy (as for the Gaussian case). But considerably larger
system sizes have to be studied to find these low-energy paths, 
because, in contrast to the Gaussian case, there are no bonds which can 
be broken at the cost of very small energies. Every broken bond costs an
energy $2J$. Indeed, by introducing possible zero bonds, i.e.,~treating a
diluted system, the effect is reduced.\cite{amoruso:03}
As we will show in the next section, the scaling of the ground-state
energy is related to the physical appearance of domain-walls,
hence the same unorthodox and strong corrections are to be expected
for the $\pm J$ case.

\section{Appearance of Domain Walls}
\label{sec:dwp}
\begin{figure}
\centerline{\epsfxsize=6.1cm 
 \epsfbox{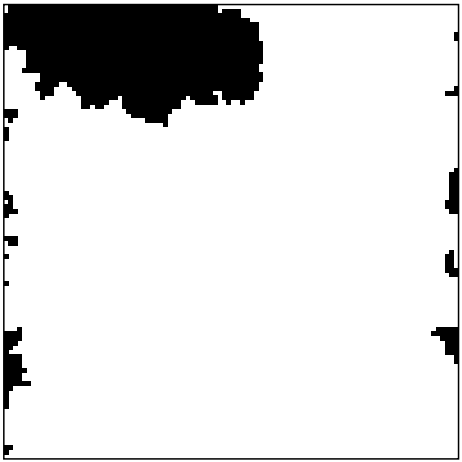}}
\caption{
Difference between the ground states of one realization ($L=100$) for free and
periodic boundary conditions in the $x$-direction. 
Bonds are changed along the vertical axes.}
\label{fig:free-periodic}
\end{figure}

We have seen, in particular for the model with a Gaussian distribution of the
interactions, that for free boundary conditions the size dependence of the
ground-state energy can be
explained by simple algebraic terms. As soon as periodic boundary conditions
are applied, a term $\sim L^{-(d-\theta)}$ appears, 
$\theta$ being the exponent describing the
size-dependence of the energy of elementary excitations such as droplets or
domain walls. In this section we want to therefore test  whether really the 
occurrence of
domain walls, induced by the boundary conditions, is responsible for this
scaling. For this purpose, we study
systems with Gaussian interactions and ffbc, calculate the ground
state, then we change the boundary conditions to pfbc (by adding one column of
random bonds wrapping around the system), recalculate the ground
state, and compare the two different ground states. In
Fig.~\ref{fig:free-periodic} the
result for one sample of size $L=100$ is shown. The result is typical: no
system-spanning domain walls are obtained, but a collection of smaller and few
larger droplet-like excitations pinned to the boundary.
We observe that quite often the larger droplets span even 
larger fractions of the systems.

To test whether system-spanning 
domain walls play a role in this framework, we perform the
following simulation: We start with ffbc, calculate the ground state,
and then switch to pfbc. Now a fraction $p$ of the bonds which wrap around the
system is chosen as {\em inverted hard bonds}. This means that they have a
very large magnitude, such that they are satisfied in any ground
state. The sign is chosen in a way such that it forces the two adjacent spins 
to take opposite relative orientations with respect to the ffbc ground
state, i.e. exactly one spin flips. 
This means, the inverted hard bonds force a domain wall into the
system at this position. 
The remaining fraction $(1-p)$ of the additional bonds is again chosen
from a Gaussian distribution with zero mean and standard deviation unity.  
The number of samples studied as a function of $L$ and $p$ are shown in 
Table \ref{tab:paramspgt0}.

\begin{table}
\caption{
Number of samples for the ground-state calculations for
the samples (Gaussian distribution of the bonds)
having first full free, then free boundary conditions in
the $y$-direction and mixed random-weak/inverted-hard boundary
conditions in the $x$-direction (with a fraction $p$ of the boundary bonds
being inverted hard), for different system sizes $L$ and values of $p$.
For the different values of $p>0$, always the same number of samples
are taken for a fixed size, except for the largest size $L=160$. 
\label{tab:paramspgt0}
}
\begin{tabular*}{\columnwidth}{@{\extracolsep{\fill}} c c l  }
\hline
\hline
$L$  & $p = 0$ & $p > 0$ \\
\hline
6   & 21000 & 1000     \\
10  & 21000 & 1000     \\
16  & 20000 &          \\
20  & 10000 & 1000     \\
30  & 20000 &          \\
40  & 20000 & 1000     \\
60  & 15000 &          \\
80  & 13000 & 1000     \\
120 & 15000 &          \\
160 & 15250 & 700-4600 \\
\hline
\hline
\end{tabular*}
\end{table}

\begin{figure}
\centerline{\epsfxsize=\columnwidth \epsfbox{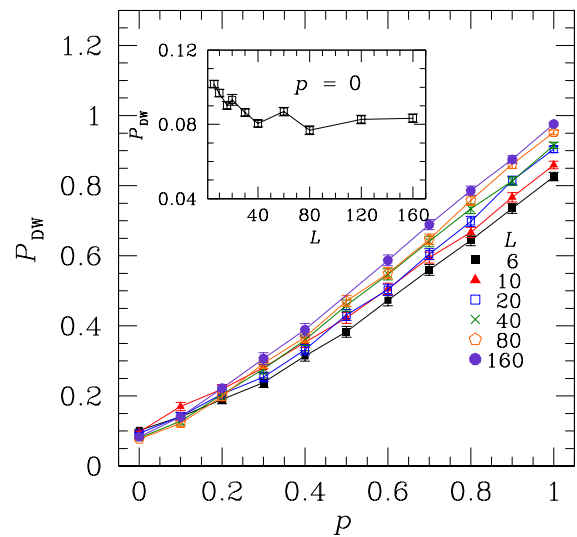}}
\vspace{-1.0cm}
\caption{(Color online)
Probability $P_{\rm DW}$
of the occurrence of a system-spanning domain wall as a function of the
fraction $p$ of inverted hard bonds wrapping around the system, for different
system sizes $L$. The lines connect the dots and are guides to the eye only. 
The inset shows $P_{\rm DW}(p=0)$ as a function of $L$. We see that for
$p = 0$ in average only $\sim$ 10\% of the domain walls are system spanning.
}
\label{fig:p_dw_fig}
\end{figure}

In Fig.~\ref{fig:p_dw_fig} the probability $P_{\rm DW}$ 
that a domain wall spans from the
top to the bottom is shown as a function of $p$. Clearly, $P_{\rm DW}$
increases with growing $p$ and reaches unity for largest system sizes when $p\to
1$, as expected. For large $p$,  $P_{\rm DW}$ grows with system
size. Near $p=0$, no such trend is visible, see also the the inset 
of Fig.~\ref{fig:p_dw_fig}. In particular, no clear crossing of
the curves can be observed, which would be an indication that a finite
fraction of inverted hard bonds is necessary to create system spanning domain
walls. Thus, system-spanning domain walls exist in the thermodynamic limit
also for $p=0$, but not many.

\begin{figure}
\centerline{\epsfxsize=\columnwidth \epsfbox{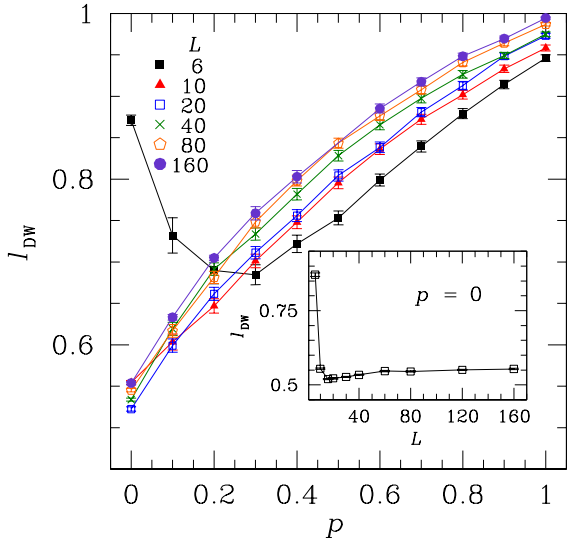}}
\vspace{-1.0cm}
\caption{(Color online)
Length $l_{\rm DW}$ (in units of $L$ along the $y$-direction) 
of the longest nontrivial domain wall as a function of  the
fraction $p$ of inverted hard bonds wrapping around the system, for different
system sizes $L$. Lines are guides to the eyes only. The inset shows $l_{\rm
  DW}(p=0)$ as a function of $L$.
}
\label{fig:l_dw_fig}
\end{figure}

To obtain a clearer picture, we also measure for each system 
the size  
of the largest nontrivial domain wall (the largest trivial domain contains
the spins which have not changed) in units of the system size $L$, when
projected onto the $y$ axis. The resulting length $l_{\rm DW}$ 
averaged over disorder is displayed in 
Fig.~\ref{fig:l_dw_fig}. On can see that the domain walls grow with
increasing fraction $p$ of the inverted hard  bonds. For $p=0$ (see 
the inset of Fig.~\ref{fig:l_dw_fig}),
$l_{\rm DW}$  seems to converge to a quite large value, larger than 0.5.
This means that the periodic boundary conditions create domain walls of order
system size, although not spanning the system. For these domain walls 
(one can call them also droplets pinned at the border) 
indeed a scaling of their energy proportional to
$L^\theta$ can be expected. Hence, there is a intuitively clear reason for the
appearance of the nontrivial scaling term, which 
we have found above when studying
the ground-state energy of two-dimensional systems.

\section{Conclusions}
\label{sec:conclusions}

We have analyzed the size-dependent energies for two-dimensional ISGs with
Gaussian and $\pm J$ interactions, up to large sizes and for
three different types of boundary conditions: free along both axes of the
sample (ffbc), periodic (ppbc) and free/periodic (pfbc).

With a simple scaling expression and three free parameters, excellent fits
can be obtained to the ffbc free boundary condition data, from the
isolated spin $L=1$ up to large sizes, $L=320$ for the Gaussian and
$L=480$ for the $\pm J$ cases. The fitting parameters can be translated into
bulk-, edge-, and corner-spin (or bond) energies. The bulk ground state
energy per spin is $e_{\infty}=-1.31479(2)$ for the Gaussian and
$e_{\infty}=-1.40197(1)$ for the $\pm J$ case. 
The former is considerably more
accurate than previous estimates, and the latter is consistent with the
high-precision published value.\cite{palmer:99} The effective edge and corner 
bond energies do not
seem to have been measured before; they are more negative than the bulk
values as might be expected because of geometrically reduced frustration.
It is remarkable that although the entire range of $L$ including $L=1$ is
used, three parameters only are sufficient for an excellent fit in both
cases.

For the periodic-periodic boundary condition geometry, a scaling 
analysis of the Gaussian system
data can be made using only parameters ($e_{\infty}$ and the stiffness
exponent $\theta$) whose values are already accurately known. The zero free-parameter ``fit'' passes through $e_{L=1}=0$
just as for the ffbc case. Thus for the Gaussian system the scaling in these
two geometries appears to be excellent, implying that any correction terms present are very small.

For the $\pm J$ system, the ppbc fit is less satisfying than for the
Gaussian case, as an acceptable fit requires the introduction of a correction to
scaling term. Even with this supplementary term included the global fit is
not as good as for the Gaussian case.

In the mixed pfbc geometry the fit to the Gaussian data needs the
introduction of a weak analytic correction term, and the extrapolation to
$L=1$ does not quite go through $e_{L=1}=0$. The scaling is thus slightly less
aesthetically pleasing than for the ffbc or ppbc geometries but follows
standard correction to scaling rules.

For the $\pm J$ system any fit to the pfbc data based on the same approach
as for the Gaussian is very poor. In this geometry, for the size
dependence of the energy as for
the directly measured domain-wall stiffness, 
there are strong deviations from scaling that do not follow the orthodox
correction to scaling behavior. As these deviations do not appear in
the Gaussian case, we associate them  with the degenerate ground state
of the $2d \pm J$ ISG, as discussed above. 
This hypothesis could be checked by measurements on other $2d$ ISG
systems with degenerate ground states. 

By studying explicitly sample by sample the same system having ffbc and 
pfbc, we have seen that indeed domain walls of order of the system size are 
created by the periodic boundary
conditions, although usually these domain walls (or pinned droplets) do not
span the whole system. The appearance of these domain walls explains the
occurrence of the $L^\theta$ term in the size dependence of the ground-state
energy for periodic boundary conditions in a quite natural way.

\begin{acknowledgments}

We would like to thank A.~P.~Young for useful comments and suggestions on a
previous version of the manuscript.
Parts of the simulations were performed at the Paderborn Center
for Parallel Computing (Germany) and on a Beowulf Cluster
at the Institut f\"ur Theoretische Physik of the Universit\"at
Magdeburg (Germany). 
We would like to thank the group of Professor Michael J\"unger at the 
University 
of Cologne for placing their spin-glass server (Ref.~\onlinecite{juenger:sg}) 
in the public
domain, and for not complaining when we crashed their mail server twice due to
large amounts of incoming ground-state configurations. In particular we would
like to thank Frauke Liers for help concerning the spin-glass server.
AKH obtained financial support from the
{\em VolkswagenStiftung} (Germany) within the program
``Nachwuchsgruppen an Universit\"aten'' and from the Institute for
Scientific Interchange (ISI) Foundation in Turin via the Complex
Systems Network of Excellence ``Exystence.'' 

\end{acknowledgments}

\bibliography{refs,comments}

\end{document}